\documentclass{aa}
\usepackage{graphicx}
\usepackage{amsmath}
\usepackage{amssymb}
\usepackage{bm}
\usepackage{url}
\usepackage[colorlinks=true,citecolor=blue,linkcolor=blue,urlcolor=blue]{hyperref}
\usepackage{xcolor}

\providecommand{\apjl}{ApJ}

\providecommand{\aap}{A\&A}

\newcommand{\dth}{\Delta\theta}

\makeatletter
\@ifundefined{proposition}{%
  \newtheorem{proposition}{Proposition}}{}
\@ifundefined{proof}{%
  \newenvironment{proof}{\par\noindent\textit{Proof.}\ }{\hfill$\square$\par\medskip}}{}
\makeatother

\begin{document}

\title{An exact and fast solution for the common overlap of three disks in mutual transits and eclipses}

\titlerunning{Common overlap of three disks}
\authorrunning{M. Hippke \& R. Heller}

\author{Michael Hippke\inst{1}
        \and
        Ren{\'e} Heller\inst{2}}

\institute{Sonneberg Observatory, Sternwartestr. 32, 96515 Sonneberg, Germany\\
           \email{michael@hippke.org}
           \and
           Max-Planck-Institut f{\"u}r Sonnensystemforschung, Justus-von-Liebig-Weg 3, 37083 G{\"o}ttingen, Germany\\
           \email{heller@mps.mpg.de}}

\date{Received ; accepted }

\abstract{The area common to three disks determines how much light is blocked during
mutual events of transiting planets, moons, and eclipsing multiple stars. We
present a new solution that removes every case distinction from the evaluation
and corrects several errors in earlier work. We prove that a non-degenerate
overlap has at most four corners, which reduces the geometry to five
topologies that a single directed walk along the boundary arcs resolves.
Green's theorem then yields the area as a sum of circular segments and a
vertex polygon. The evaluation needs one arctangent per arc, suffers no
subtractive cancellation, and handles arcs longer than $\pi$ without a special
case. We verify the solution for all cases and show that the algorithm is
componentwise backward stable. Five public packages that compute the same area
fail on some of our test configurations, either through undefined values,
missing topologies, or lost precision. Our complete computation takes 45\,ns
per configuration on one CPU core, which is four to several thousand times
faster than the public packages. The same decomposition also delivers the
centroid of the overlap in closed form. An open-source implementation is
available at \protect\url{https://github.com/hippke/reuleaux}.}


\maketitle
\nolinenumbers

\section{Introduction} \label{sec:intro}

When two bodies transit a star at the same time and overlap on the sky, the
stellar light they block equals the area of the union of their shadows on the
stellar disk. We represent the star, a planet, and its moon by three disks. The
union then follows from the pairwise lenses and one further quantity, the area
of the common overlap of all three disks. Mutual events of this kind constrain
the orbits of exomoons \citep{CabreraSchneider2007, 2011ApJ...743...97T, Kipping2011,
HippkeHeller2022, Kalman2023}, planet--planet occultations \citep{Luger2017},
and multiply eclipsing stellar systems \citep{Short2018}. The occultation by a
single body has an analytic solution \citep{MandelAgol2002, Agol2020,
Luger2019}, and the geometry of three disks underlies every analytic treatment
of the mutual case.

\citet{Fewell2006} gave the first closed form for the common overlap of three
circles. He split the circular triangle into the straight triangle on its three
chords, which he evaluated with Heron's formula, and three circular segments,
and he sorted the configurations into nine cases of which only one is a
circular triangle. For a segment that exceeds half of its disk, his Eq.~16
retains the arcsine of the chord, $r^2\arcsin(c/2r)$, which measures the minor
arc, and omits the term $\pi r^2$. \citet{Kipping2011} transcribed Fewell's
construction to the star--planet--moon problem in the algorithm LUNA. He
organised the configurations into 27 principal cases with subcases and thereby
inherited the error in his Eq.~37 (case 14.1b). \citet{GordonAgol2022} found a
discrepancy between LUNA and an exact computation near the inner edge of the
stellar limb and published the corrected segment,
$r^2[\pi-\arcsin(c/2r)] + (c/4)\sqrt{4r^2-c^2}$, in their Appendix~B. The root
cause is structural. A chord length fixes an arc only up to the choice between
the arc and its complement, so every formula built on chord lengths needs an
additional geometric test to make that choice.

The exact limb-darkened light curve of two mutually transiting bodies follows
from Green's theorem, which converts the flux integral over the visible stellar
disk into line integrals along its boundary arcs \citep{Pal2012}.
\citet{GordonAgol2022} implemented this approach in the code \texttt{gefera}
with a tree of 16 geometric cases.
\citet{Short2018} integrate the line integrals numerically, \citet{Luger2017}
discretise the star into annuli, and \citet{Kalman2023} combine pixel
classification with Gauss--Legendre quadrature in the Transit and Light Curve
Modeller \citep{Csizmadia2020}. Outside astronomy, \citet{Librino2014}
computed intersections of many circles by inclusion--exclusion over a trellis
of pairwise terms and confirmed Fewell's construction as the reference for
three circles. Area-proportional Euler and Venn diagrams require the same
quantity. The packages \texttt{eulerr} \citep{eulerr}, \texttt{venn.js}
\citep{vennjs}, and \texttt{matplotlib-venn} \citep{mplvenn} compute
intersections of circles from their crossing points, polygon areas, and
circular segments, whereas general geometry libraries such as Shapely
\citep{shapely} on GEOS \citep{geos} approximate disks by polygons. LUNA is
not public.

All of these approaches rest on the area of a two-disk lens. Its textbook form
\citep{Weisstein} is
\begin{equation}
\begin{split}
L ={}& r_1^2\arccos\frac{d^2+r_1^2-r_2^2}{2dr_1}\\
&+ r_2^2\arccos\frac{d^2+r_2^2-r_1^2}{2dr_2} - \tfrac12\sqrt{\Pi},
\end{split}
\label{eq:lens}
\end{equation}
where $\Pi=(-d+r_1+r_2)(d+r_1-r_2)(d-r_1+r_2)(d+r_1+r_2)$. This expression
subtracts quantities of nearly equal size when the lens is thin, and the
arccosine of an argument $1-\epsilon$ amplifies the rounding error of the
argument by $\epsilon^{-1/2}$. For a small disk that barely enters a large
one, as for a moon at stellar ingress, Equation~(\ref{eq:lens}) loses every
significant digit (Section~\ref{sec:accuracy}). The same cancellation affects
the intersection points computed as $h^2=r_1^2-a^2$, and these enter all
vertex-based formulas.

In this paper we present a solution that is free of case-dependent formulas,
correct for arcs of any length, backward stable in floating-point arithmetic,
and as fast as the closed form of a single topology. We derive the method in
Section~\ref{sec:method}, quantify its accuracy and cost in
Section~\ref{sec:results}, and discuss its scope in
Section~\ref{sec:discussion}.

\section{Method} \label{sec:method}

\subsection{Statement of AI use} \label{sec:ai}

The solution presented in this paper was derived with the large language model Claude Opus 5.5 (Anthropic) in September and October 2026.
The authors began with general questions to Claude and then challenged the proposed solutions with increasingly specific requests, until they could follow every step of the derivation.
The solution in its present form results from these iterations and not from a single answer of the model.
Claude also wrote the software, including our implementation and the code that compares the accuracy and speed of all methods.
It also assisted in drafting the text and in checking its spelling and grammar.
No other artificial intelligence (AI) tool was used.

\subsection{Topology} \label{sec:topology}

Let $D_k$ denote the closed disk of centre $\mathbf{o}_k$ and radius $r_k$
bounded by the circle $C_k$, $k=0,1,2$, and let $R=D_0\cap D_1\cap D_2$. We
first remove every disk $D_j$ that contains another disk $D_i$, because then
$D_i\cap D_j=D_i$. One remaining disk gives $A=\pi r^2$, and two remaining
disks give the lens of that pair. The overlap vanishes, $A=0$, as soon as one
pair is disjoint. In all other configurations every pair of circles crosses at
two points. A crossing of $C_i$ and $C_j$ that lies strictly inside $D_k$ is a
vertex of $R$, and we denote the number of vertices by $V$.

\begin{proposition}
For configurations without a crossing on the third circle, $V\in\{0,2,3,4\}$.
The overlap is empty for $V=0$, the lens of one pair for $V=2$, a circular
triangle with one vertex from each pair for $V=3$, and a circular
quadrilateral whose four vertices lie on one circle for $V=4$.
\end{proposition}

\begin{proof}
The set $R$ is convex. Unless $R$ is a disk or has an empty interior, its
boundary is a simple closed curve composed of $a_k$ arcs of each $C_k$, joined
at the vertices, so that $V=a_0+a_1+a_2$. An arc of $\partial R$ on $C_i$ is a
component of $(C_i\cap D_j)\cap(C_i\cap D_k)$, the intersection of two proper
arcs of $C_i$. Hence $a_i\le2$, and $a_i=2$ holds exactly when the two arcs
cover $C_i$, that is, when $C_i\subset D_j\cup D_k$. The union $D_j\cup D_k$ is
simply connected because $D_j\cap D_k\neq\emptyset$, and by the Jordan curve
theorem it then contains $D_i$. A crossing $\mathbf{q}$ of $C_j$ and $C_k$ lies
on the boundary of $D_j\cup D_k$ and therefore cannot lie in the interior of
$D_i$, so the pair $(j,k)$ contributes no vertex. Each vertex on $C_j$ then
stems from the pair $(i,j)$ and each on $C_k$ from $(i,k)$, which gives
$a_j=a_k=1$ and $V=4$. In all other cases $a_k\le1$ for every $k$, and hence
$V\le3$. For $V=3$ every circle carries one arc and two vertices. Two vertices
from the same pair would leave the third circle without an arc, so each pair
contributes exactly one. For $V=2$ the boundary consists of two arcs of the
circles of one pair. A boundary without vertices would be a full circle
$C_i\subset D_j\cap D_k$, which the containment reduction has removed.
Therefore $V=0$ implies $R=\emptyset$, and $V=1$ does not occur.
\end{proof}

\begin{figure*}
\includegraphics[width=\textwidth]{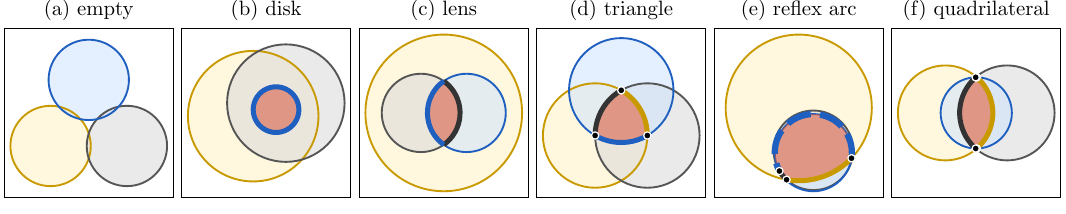}
\caption{The topologies of the common overlap (red) of three disks after
containment reduction. Vertices are marked by dots, boundary arcs are
drawn in the colour of their circle, and the dashed arc in panel (e)
exceeds $\pi$. Panel (a) shows three pairwise intersecting disks without a
common point.}
\label{fig:taxonomy}
\end{figure*}

The proposition thus replaces the 9, 27, and 16 cases of \citet{Fewell2006},
\citet{Kipping2011}, and \citet{GordonAgol2022} by a simple count, and Figure
\ref{fig:taxonomy} shows the five resulting topologies.

\subsection{Boundary walk} \label{sec:walk}

A counter-clockwise traversal of $\partial R$ runs counter-clockwise around the
centre of each of its arcs, since $R$ lies inside every disk. An arc of $C_s$
between consecutive crossings on $C_s$ has constant membership in the two other
disks and belongs to $\partial R$ exactly when its midpoint lies inside both.
From every vertex $\mathbf{u}$ we therefore select the unique counter-clockwise
arc to another vertex $\mathbf{v}$ on a shared circle $C_s$ whose open span
contains no other crossing of $C_s$ and whose midpoint lies strictly inside the
two other disks. We write $\mathbf{u}'=\mathbf{u}-\mathbf{o}_s$ and
$\mathbf{v}'=\mathbf{v}-\mathbf{o}_s$. The span test then reduces to signs of
cross products, and the midpoint is
$\mathbf{o}_s\pm r_s(\mathbf{u}'+\mathbf{v}')/|\mathbf{u}'+\mathbf{v}'|$,
with the positive sign for $\mathbf{u}'\times\mathbf{v}'>0$. The successors
form a single cycle through all vertices, and any other outcome identifies a
degenerate input. For $V=3$ each circle carries exactly two vertices, the arcs
follow the fixed adjacency of the three pairs, and the midpoint test alone
fixes their orientation. For $V=4$ the two arcs on the circle that carries all
vertices join angularly adjacent vertices in counter-clockwise order. We
evaluate both cases without enumeration.

\subsection{Area} \label{sec:area}

\begin{figure*}
\includegraphics[width=\textwidth]{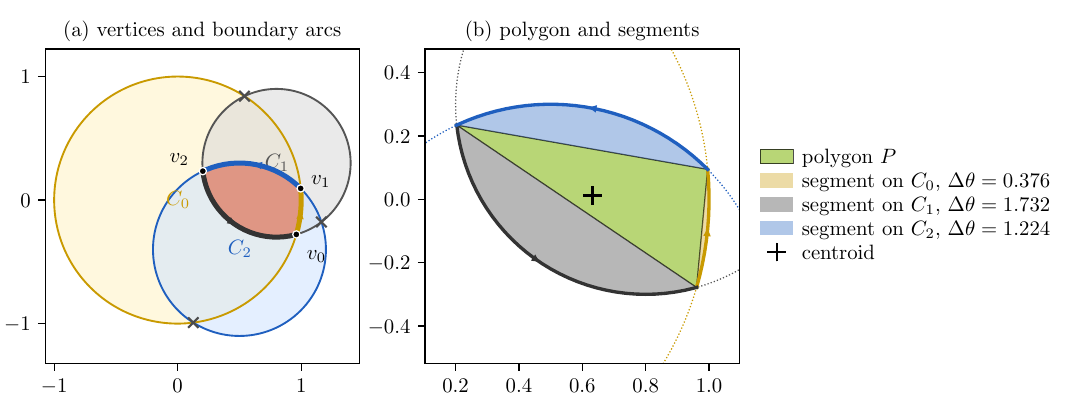}
\caption{(a) Vertices of the overlap (dots), crossings outside the third
disk (crosses) and the directed boundary arcs. (b) Decomposition of
Equation~(\ref{eq:area}) into the vertex polygon and one circular segment per
arc, with the centroid of Equation~(\ref{eq:centroid}).}
\label{fig:method}
\end{figure*}

Green's theorem gives $A=\frac12\oint(x\,dy-y\,dx)$. We parametrise the arc of
$C_s$ from $\mathbf{u}$ to $\mathbf{v}$ as $\mathbf{o}_s+r_s(\cos t,\sin t)$.
The identities $r_s(\sin\beta-\sin\alpha)=v_y-u_y$ and
$r_s(\cos\beta-\cos\alpha)=v_x-u_x$ then remove all trigonometric functions of
the end angles, and the arc contributes
\begin{equation}
A_{\rm arc}=\tfrac12\left[r_s^2\dth+o_{s,x}(v_y-u_y)-o_{s,y}(v_x-u_x)\right],
\label{eq:arc}
\end{equation}
with
\begin{equation}
\dth=\operatorname{atan2}\!\left(\mathbf{u}'\times\mathbf{v}',\,\mathbf{u}'\cdot\mathbf{v}'\right)\in(0,2\pi].
\label{eq:dtheta}
\end{equation}
Equation~(\ref{eq:dtheta}) determines the arc from the geometry rather than
from a chord length, so arcs longer than $\pi$ need no separate treatment. Each
term of Equation~(\ref{eq:arc}) equals the circular segment of the arc plus the
signed triangle $(\mathbf{o}_s,\mathbf{u},\mathbf{v})$, and the triangle terms
combine around the cycle into the polygon of the vertices. The area is
therefore
\begin{equation}
A=\sum_{\rm arcs}\frac{r_s^2}{2}\left(\dth-\sin\dth\right)
+\frac12\sum_{\rm edges}(\mathbf{u}-\mathbf{p})\times(\mathbf{v}-\mathbf{p}),
\label{eq:area}
\end{equation}
where $\mathbf{p}$ denotes the first vertex. Equation~(\ref{eq:area})
generalises the decomposition of \citet{Fewell2006} to every topology and every
arc length, and we evaluate it in a form that contains no subtractive
cancellation beyond that inherent in the data. For $\dth<0.5$ we compute the
segment from the Taylor series
$\dth-\sin\dth=\sum_{n\ge1}(-1)^{n+1}\dth^{2n+1}/(2n+1)!$, truncated after
nine terms. For $\dth\ge0.5$ we use $r_s^2\sin\dth=\mathbf{u}'\times\mathbf{v}'$,
which requires no further transcendental function. We accumulate the polygon
relative to $\mathbf{p}$, so that its rounding error scales with the size of
$R$ and not with the coordinates, and we translate the input to the centre of
the smallest disk. Figure~\ref{fig:method} illustrates the walk and the
decomposition.

\subsection{Crossings and lenses} \label{sec:crossings}

The crossings of $C_1$ and $C_2$ at distance $d$ lie at distance
$a_1=[(d-r_2)(d+r_2)+r_1^2]/2d$ from $\mathbf{o}_1$ along the line of centres
and at distance $h$ from it. Here $4d^2h^2$ is sixteen times the squared area
of the triangle with sides $(r_1,r_2,d)$. We evaluate this quantity with the
ordering and bracketing of \citet{Kahan2014}. For the sorted sides
$\alpha\ge\beta\ge\gamma$ it reads
\begin{equation}
4d^2h^2=[\alpha+(\beta+\gamma)][\gamma-(\alpha-\beta)][\gamma+(\alpha-\beta)][\alpha+(\beta-\gamma)],
\label{eq:kahan}
\end{equation}
which retains full relative accuracy for needle-like triangles, that is, for
nearly tangent circles. The lens then follows from two segments with central
angles $2\operatorname{atan2}(h,a_1)$ and $2\operatorname{atan2}(h,a_2)$, which
replaces Equation~(\ref{eq:lens}).

\subsection{Degenerate input} \label{sec:degenerate}

We resolve tangencies and containments within $10^{-12}$ of the largest radius
by the reduction of Section~\ref{sec:topology}. A crossing within $10^{-9}$ of
the third circle, a triple point, admits no reliable inside test. In this case
we evaluate the solution with all radii shifted by $\pm10^{-8}$ of the largest
radius and average the two results. For a circle through both corners of a lens
the averaged area agrees with the exact lens to $10^{-15}$, and for three
circles through one point it equals $7\times10^{-18}$.

\subsection{Centroid} \label{sec:centroid}

The decomposition of Equation~(\ref{eq:area}) also yields the first moments of
$R$. The centroid of a segment with half-angle $\varphi=\dth/2$ lies on its
axis, beyond the chord midpoint by
\begin{equation}
\begin{split}
e&=r\left[\frac{4\sin^3\varphi}{3(2\varphi-\sin2\varphi)}-\cos\varphi\right]\\
&= r\left(\frac{\varphi^2}{5}-\frac{13\,\varphi^4}{1050}+\ldots\right),
\end{split}
\label{eq:centroid}
\end{equation}
and for $\varphi<0.5$ we evaluate the series, which we derived in exact
rational arithmetic to order $\varphi^{20}$. Together with the triangle fan of
the polygon, this gives the centroid of the overlap to within
$1.7\times10^{-12}$ of the smallest radius against an independent quadrature.
The centroid provides the first-order correction for the variation of the
stellar intensity across the overlap, which constant-intensity treatments of
mutual events omit.

\section{Results} \label{sec:results}

\subsection{Accuracy} \label{sec:accuracy}

We measure accuracy against references computed with 50 significant digits in
an arbitrary-precision library \citep{mpmath}. The reference re-evaluates every
vertex, angle, segment, and polygon of Equation~(\ref{eq:area}) in extended
precision and takes the topology from the double-precision classification.
That classification agrees to $3.2\times10^{-14}$ of the smallest disk, over
$20{,}000$ configurations, with an independent adaptive quadrature of the
chord length $\min_k(y_k+s_k)-\max_k(y_k-s_k)$, which contains no geometric
case logic. We sample the problem with five families of configurations, namely
general disks, disks of similar radius with strong overlap, a hierarchical
star--planet--moon geometry with radii $1$, $0.02$--$0.2$, and $0.005$--$0.1$,
circular triangles produced by clipping a lens corner to a depth of $10^{-7}$
to $0.5$ times the smallest radius, and thin lenses of a small disk at the rim
of a large one.

\begin{figure*}
\includegraphics[width=\textwidth]{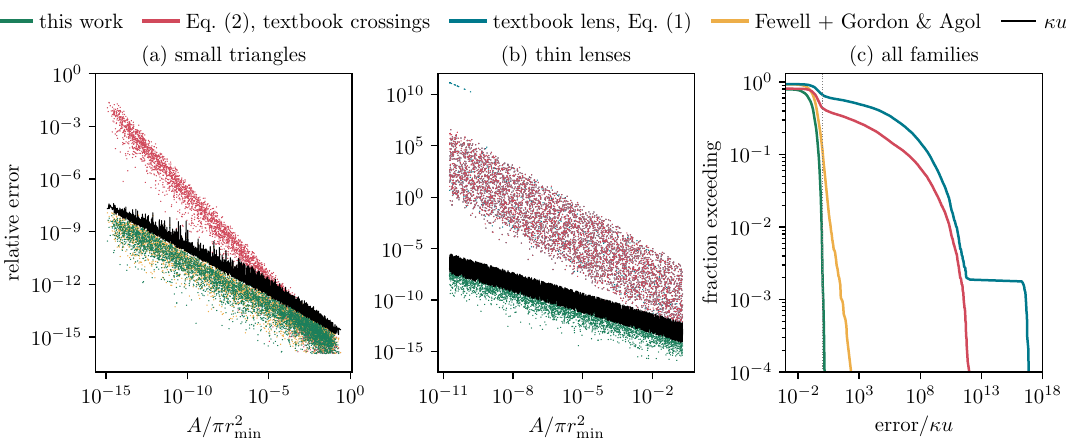}
\caption{Accuracy against 50-digit references. (a) Relative error of small
circular triangles versus their area in units of the smallest disk. (b)
Relative error of thin lenses. The black curve marks $\kappa u$. (c)
Fraction of all $21{,}568$ configurations whose error exceeds a given
multiple of $\kappa u$.}
\label{fig:accuracy}
\end{figure*}

The conditioning of the problem limits the attainable accuracy. We compute the
componentwise relative condition number
$\kappa=\sum_i|p_i\,\partial A/\partial p_i|/A$ over the nine input
parameters by differentiation in extended precision. An algorithm whose
error does not exceed a small multiple of $\kappa u$, with
$u=2^{-53}$ the unit roundoff, returns the exact area of a configuration
perturbed by a few units of roundoff in its data and is therefore backward
stable in the componentwise sense \citep{Higham2002}. Table~\ref{tab:accuracy}
and Figure~\ref{fig:accuracy} compare our method with the direct evaluation of
Equation~(\ref{eq:arc}) using the textbook crossings and
Equation~(\ref{eq:lens}), with the corrected closed form of
\citet{Fewell2006} and \citet{GordonAgol2022} for circular triangles, and
with Equation~(\ref{eq:lens}) for lenses.

\begin{table*}
\caption{Maximum relative error and maximum error in units of $\kappa u$.}
\label{tab:accuracy}
\centering\small
\begin{tabular}{lrcccc}
\hline\hline
Family & $N$ & This work & Eq.~(\ref{eq:arc}), textbook & Eq.~(\ref{eq:lens}) & Fewell + G\&A \\
\hline
general      & 2155 & $1.9\times10^{-13}$ / 1.1 & $2.0\times10^{-9}$ / $7.3\times10^{2}$ & $2.0\times10^{-9}$ / $7.3\times10^{2}$ & $1.5\times10^{-13}$ / 30 \\
similar      & 4995 & $1.4\times10^{-14}$ / 2.0 & $2.2\times10^{-13}$ / 5.8 & $4.2\times10^{-13}$ / 9.0 & $4.0\times10^{-14}$ / $1.3\times10^{2}$ \\
hierarchical & 4418 & $1.5\times10^{-12}$ / 1.3 & $2.1\times10^{-7}$ / $9.1\times10^{4}$ & $2.1\times10^{-7}$ / $9.1\times10^{4}$ & $3.4\times10^{-12}$ / $2.1\times10^{2}$ \\
corner       & 5000 & $9.9\times10^{-9}$ / 1.0 & $2.3\times10^{-2}$ / $1.1\times10^{6}$ & -- & $1.2\times10^{-8}$ / 1.4 \\
sliver lens  & 5000 & $1.5\times10^{-6}$ / 0.9 & $3.6\times10^{6}$ / $1.3\times10^{12}$ & $1.4\times10^{11}$ / $7.8\times10^{16}$ & -- \\
\hline
\end{tabular}
\tablefoot{$N$ is the number of configurations with non-zero area.
Each entry gives the maximum relative error followed by the maximum of
the error divided by $\kappa u$. Equation~(\ref{eq:lens}) applies to
lens results, and the closed form of \citet{Fewell2006} with the correction
of \citet{GordonAgol2022} to circular triangles.}
\end{table*}

The error of our method never exceeds $2.01\,\kappa u$ in any of the
$21{,}568$ configurations, and its median is $0.12\,\kappa u$. The relative
errors of $10^{-8}$ for the smallest triangles and $10^{-6}$ for the thinnest
lenses are those of the data, because a relative perturbation of $10^{-16}$ in
the coordinates changes these areas by the same amounts. The direct evaluation
of Equation~(\ref{eq:arc}) with the textbook crossings exceeds this limit by up
to six orders of magnitude for small triangles, and Equation~(\ref{eq:lens})
does so by up to 16 orders of magnitude for thin lenses, where it returns areas
wrong by factors up to $10^{11}$. The corrected closed form of Fewell remains
within $2\times10^{2}\,\kappa u$ but covers circular triangles only.

\subsection{Arcs longer than $\pi$}

\begin{figure*}
\includegraphics[width=\textwidth]{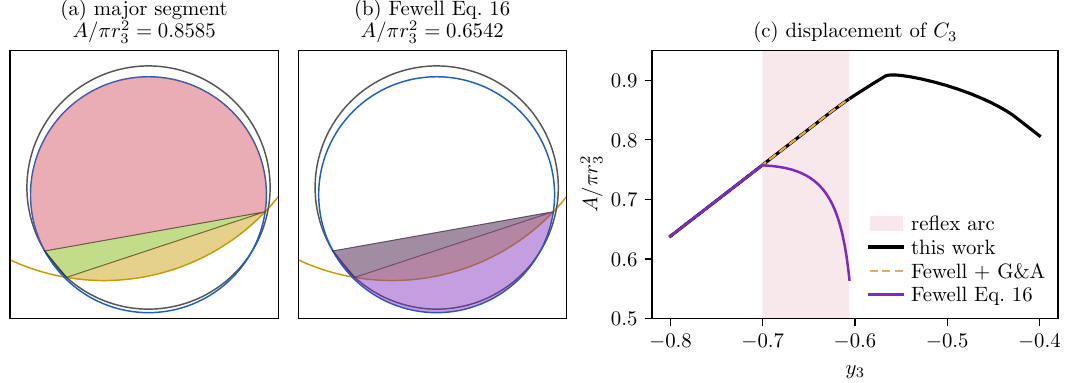}
\caption{A circular triangle with an arc of $3.78$\,rad on the smallest
circle $C_3$. (a) Heron triangle and segments with the major segment on
$C_3$, which gives the exact area. (b) The segment of Fewell's Eq.~16,
which takes the minor arc. (c) Overlap area as $C_3$ moves along $y$.
The shaded interval contains a reflex arc.}
\label{fig:reflex}
\end{figure*}

Among the circular triangles of Table~\ref{tab:accuracy}, 101 contain an arc
longer than $\pi$. On these configurations the original Eq.~16 of
\citet{Fewell2006}, which \citet{Kipping2011} adopted as Eq.~37, errs by up to
66\% of the area. Figure~\ref{fig:reflex} shows one example, in which the
original formula reduces the overlap from $0.859$ to $0.654$ of the smallest
disk. A displacement of the smallest disk shows that the original formula is
continuous at the transition $\dth=\pi$ and diverges beyond it, which explains
why the error escaped detection. Our method reproduces the corrected closed
form to $2.4\times10^{-12}$ of the smallest disk over all 4462 circular
triangles and never tests whether an arc is reflex.

\subsection{Public software} \label{sec:software}

We evaluated five public packages on the $40{,}000$ configurations of
Section~\ref{sec:accuracy}, of which $21{,}568$ have a non-zero overlap and
$18{,}432$ an empty one. The packages are \texttt{gefera} \citep[commit
e27e832,][]{GordonAgol2022}, \texttt{eulerr} 7.1.0 \citep{eulerr},
\texttt{venn.js} \citep{vennjs}, \texttt{matplotlib-venn} 1.1.2
\citep{mplvenn}, and Shapely 2.1.2 with GEOS 3.14.1 \citep{shapely, geos} at 8,
64, and 512 polygon segments per quarter circle. We compiled \texttt{gefera}
from its Fortran source with its own flags and, with identical results, with
\texttt{-O2}. The code returns the flux of a star occulted by two bodies. For a
uniform source this flux equals the area of the union of the two shadows on the
star, $|S\cap(P\cup M)|$, and we compare that quantity directly, which spares
\texttt{gefera} the cancellation of an inclusion--exclusion. We count an
undefined result, an exception, a relative error above $10^{-3}$ for a
non-empty overlap, or a non-zero result exceeding $10^{-9}$ of the smallest
disk for an empty one as a gross failure. Table~\ref{tab:software} summarises
the results.

\begin{table*}
\caption{Public software on $40{,}000$ configurations.}
\label{tab:software}
\centering\small
\begin{tabular}{lcccc}
\hline\hline
Code & Undefined or exception & Wrong by $>10^{-3}$ & Max.\ error$/\kappa u$ & Median error$/\kappa u$ \\
\hline
This work                         & 0    & 0     & 2.0                 & 0.12 \\
\texttt{gefera}, union area       & 31   & 90    & --             & -- \\
\texttt{eulerr} 7.1.0             & 0    & 766   & $9.7\times10^{13}$  & 2.7 \\
\texttt{venn.js}                  & 0    & 3689  & $7.0\times10^{13}$  & 2.6 \\
\texttt{matplotlib-venn} 1.1.2    & 3746 & 1903  & $4.5\times10^{15}$  & 10 \\
Shapely, 8 segments               & 0    & 21568 & $7.0\times10^{14}$  & $1.1\times10^{13}$ \\
Shapely, 64 segments              & 0    & 11022 & $4.6\times10^{11}$  & $2.1\times10^{11}$ \\
Shapely, 512 segments             & 0    & 7294  & $7.1\times10^{9}$   & $3.5\times10^{9}$ \\
\hline
\end{tabular}
\tablefoot{Errors are given relative to 50-digit references, and
$\kappa u$ is defined as in Table~\ref{tab:accuracy}. For \texttt{gefera} we
compare the union of the two shadows on the star, whose condition number
differs from that of the triple overlap. Its maximum relative error is 0.95.
Its 31 undefined results occur for a planet inside the moon, which lies outside the domain of \texttt{gefera} (Section~\ref{sec:software}).}
\end{table*}

\texttt{gefera} computes the union with a median relative error of
$5\times10^{-16}$ and agrees to $2.3\times10^{-10}$ with the reference in all
hierarchical configurations. It returns undefined values, however, in all 31
configurations in which the disk of the planet lies inside that of the moon.
This behaviour is expected, because \texttt{gefera} is designed for physical configurations, in which the moon is not larger than the planet.
Within its domain, which includes bodies larger than the star \citep{GordonAgol2022}, \texttt{gefera} assigns wrong areas, with errors up to 95\%, to 90 circular
quadrilaterals, 88 of which involve a body larger than the star. It also
returns undefined values when a crossing of planet and moon lies exactly on the
limb and when $b+r=1$ for a single body. \texttt{eulerr} agrees with the
reference in general configurations but fails on 310 small triangles and 448
thin lenses, and on 8 hierarchical configurations with bodies of radius $0.02$
and $0.006$, where it returns either zero or the full smallest disk in place of
overlaps of 24 to 94\% of that disk. \texttt{venn.js} fails on 1473 small
triangles and 2216 thin lenses, with relative errors up to $1.6\times10^{6}$.
It evaluates circular segments from their width with an arccosine of $1-w/r$
and tests containment with an absolute tolerance of $10^{-10}$.
\texttt{matplotlib-venn} raises an exception in 3746 configurations in which
one disk lies inside another, because its region algebra excludes holes. It
returns zero in 886 configurations in which the smallest disk lies inside both
others and fails on 1017 overlaps smaller than $5\times10^{-7}$ of the
smallest disk. Shapely is limited by its polygon resolution. With 512 segments
per quarter circle its median relative error is $1.9\times10^{-5}$, and small
and thin overlaps are lost entirely.

\subsection{Speed} \label{sec:speed}

We implemented the method in Python compiled with Numba \citep{Lam2015} and
timed batches of $2\times10^4$ configurations per topology on one core of an
Intel Core Ultra 5 226V, taking the best of nine repetitions. The cost per
configuration, including classification, translation, and evaluation, depends
on the topology. A representative mixture of all families costs 45\,ns and is
dominated by the early exits of empty and contained configurations at
21--31\,ns, while circular triangles cost 0.17\,$\mu$s and quadrilaterals
0.26\,$\mu$s. The specialised walks of Section~\ref{sec:walk} halve the cost
relative to the general walk, which costs 117--427\,ns. The stable evaluation
of Equations (\ref{eq:area})--(\ref{eq:kahan}) costs the same as the direct
evaluation on the mixture and at most 41\% more on lenses. The closed form of
\citet{Fewell2006} with the correction of \citet{GordonAgol2022} costs
0.10\,$\mu$s for a circular triangle when the topology is known in advance, but
it requires a case tree for all other configurations.

\begin{figure}
\includegraphics[width=\columnwidth]{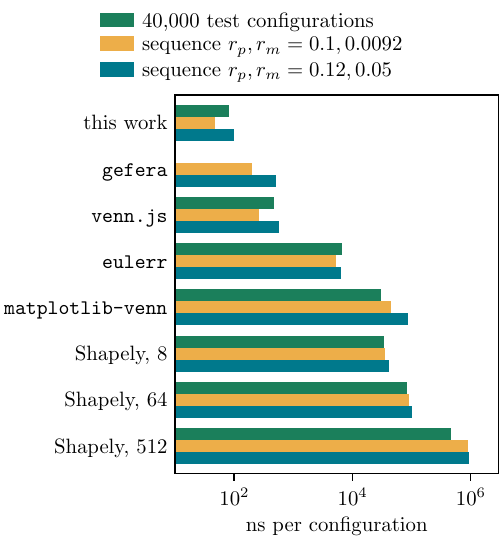}
\caption{Cost per configuration on one CPU core (Table~\ref{tab:speed}),
logarithmic scale, for the $40{,}000$ test configurations and two
light-curve sequences of a planet and a moon with fixed radii.
\texttt{gefera} accepts one pair of radii per call and is timed on the
sequences only. Shapely is shown at 8, 64, and 512 segments per quarter
circle.}
\label{fig:speed}
\end{figure}

Table~\ref{tab:speed} and Figure~\ref{fig:speed} compare the public packages on
the same core, each through its native interface, on the $40{,}000$ test
configurations and on two light-curve sequences of $2\times10^{5}$ positions of
a planet and a moon on the stellar disk with fixed radii. Our method is 4--5
times faster than \texttt{gefera}, 6 times faster than \texttt{venn.js}, and 80
to 5700 times faster than \texttt{eulerr}, \texttt{matplotlib-venn}, and
Shapely.

\begin{table*}
\caption{Cost per configuration in ns.}
\label{tab:speed}
\centering\small
\begin{tabular}{lccc}
\hline\hline
Code & Test set & $r_p,r_m=0.1,0.0092$ & $0.12,0.05$ \\
\hline
This work (Numba)            & 82      & 47      & 99 \\
\texttt{gefera} (Fortran)    & -- & 202     & 514 \\
\texttt{venn.js} (Node.js)   & 465     & 265     & 573 \\
\texttt{eulerr} (R)          & 6825    & 5350    & 6550 \\
\texttt{matplotlib-venn}     & 30074   & 45667   & 86749 \\
Shapely, 8 segments          & 34761   & 35978   & 41859 \\
Shapely, 64 segments         & 83172   & 92478   & 103402 \\
Shapely, 512 segments        & 467650  & 907037  & 949028 \\
\hline
\end{tabular}
\tablefoot{\texttt{gefera} accepts one pair of radii per call and is
therefore timed on the sequences only. It also evaluates its limb-darkening
integrals for a uniform source. The \texttt{eulerr} times include 1.8\,$\mu$s
of R loop overhead per configuration, and the \texttt{venn.js} times include the
construction of its circle objects. The test set contains more triangles
than the mixture quoted in the text.}
\end{table*}

\section{Discussion} \label{sec:discussion}

Our method separates topology from evaluation. The proposition reduces the
topology to the number of vertices, and the walk determines the boundary
without any formula that depends on the configuration. A single expression then
evaluates every topology. This structure removes the failure modes of earlier
work at their origin. The reflex error of \citet{Fewell2006} and
\citet{Kipping2011} cannot occur because no arc is inferred from a chord, and
the precision loss of Equation~(\ref{eq:lens}) cannot occur because we form
neither an arccosine of an argument near unity nor a difference of nearly equal
areas. The resulting algorithm is backward stable in the componentwise sense
over all tested geometries, including the near-tangent configurations that
arise at every ingress and egress of a mutual event. The failures of the public
packages in Section~\ref{sec:software} fall into the same two classes. The
first class comprises case trees that omit topologies, such as
quadrilaterals in \texttt{gefera} or holes in \texttt{matplotlib-venn}. The
second class is the loss of the small and thin overlaps of ingress and egress,
which stems from the segment formula and the absolute tolerance in
\texttt{venn.js} and also affects \texttt{eulerr} and \texttt{matplotlib-venn}.

The cost of the complete computation amounts to a few square roots and one
arctangent per boundary arc. In a photodynamical model the overlapping
configurations form a small fraction of all samples, and most evaluations end
in the early exits at 21--31\,ns. The method therefore removes the three-disk
geometry as a computational concern from light-curve models, and its error is
bounded by the conditioning of the data.

To first order in the gradient of the stellar intensity, the area and the
centroid of the overlap determine the light blocked twice during a mutual
event. The exact limb-darkened flux requires the line integrals of
\citet{Pal2012} and \citet{GordonAgol2022} along the same boundary arcs. Our
walk of Section~\ref{sec:walk} supplies these arcs, their end points, and their
orientation for every topology, and the stable crossings of
Equation~(\ref{eq:kahan}) supply accurate end points for the elliptic
integrals. The boundary walk thus serves as a robust front end for exact
mutual-transit photometry. The proposition and the walk also extend without
change to disks of any relative size, including bodies larger than the star in
stellar triples.

\section{Conclusions} \label{sec:conclusion}

We have presented an exact solution for the area of the common overlap of three
disks that is complete, backward stable, and fast. A proof that the overlap has
at most four corners reduces the geometry to five topologies, which one
boundary walk resolves. Green's theorem, evaluated as circular segments with
series for thin segments, a vertex polygon relative to its first vertex, and
crossings from Kahan's stable triangle area, yields errors below
$2.01\,\kappa u$ for all $21{,}568$ tested configurations. The classical
formulas err by factors up to $10^{16}$ of this bound, or for reflex arcs by up
to 66\% of the area. Five public packages, including the mutual-transit code
\texttt{gefera}, fail on between 121 and $21{,}568$ of $40{,}000$
configurations, whereas our method fails on none. The complete computation
costs 45\,ns per configuration on average and 0.26\,$\mu$s in the most complex
topology, which is four to several thousand times less than the public
packages. The same decomposition also gives the centroid of the overlap in
closed form. An open-source implementation is available at
\url{https://github.com/hippke/reuleaux}.

\begin{acknowledgements}
R.~Heller acknowledges support from the German Aerospace Agency (Deutsches Zentrum f{\"u}r Luft- und Raumfahrt) under PLATO Data Center grants 50OO1501 and 50OP1902.
This research made use of NumPy \citep{Harris2020}, Numba \citep{Lam2015},
mpmath \citep{mpmath}, Matplotlib \citep{Hunter2007}, \texttt{gefera}
\citep{GordonAgol2022}, \texttt{eulerr} \citep{eulerr}, \texttt{venn.js}
\citep{vennjs}, \texttt{matplotlib-venn} \citep{mplvenn}, Shapely
\citep{shapely}, and GEOS \citep{geos}.
\end{acknowledgements}

\bibliographystyle{aa}
\bibliography{references}

@techreport{Fewell2006,
  author      = {{Fewell}, M.~P.},
  title       = {Area of Common Overlap of Three Circles},
  institution = {Defence Science and Technology Organisation, Australia},
  number      = {DSTO-TN-0722},
  year        = {2006}
}

@article{Kipping2011,
  author  = {{Kipping}, D.~M.},
  title   = {{LUNA}: an algorithm for generating dynamic planet-moon transits},
  journal = {\mnras},
  volume  = {416},
  pages   = {689},
  year    = {2011},
  doi     = {10.1111/j.1365-2966.2011.19086.x}
}

@article{GordonAgol2022,
  author  = {{Gordon}, T.~A. and {Agol}, E.},
  title   = {Analytic light curve for mutual transits of two bodies across a limb-darkened star},
  journal = {\aj},
  volume  = {164},
  pages   = {111},
  year    = {2022},
  doi     = {10.3847/1538-3881/ac82b1},
  eprint  = {2207.06024}
}

@ARTICLE{2011ApJ...743...97T,
       author = {{Tusnski}, Luis Ricardo M. and {Valio}, Adriana},
        title = "{Transit Model of Planets with Moon and Ring Systems}",
      journal = {\apj},
         year = 2011,
        month = dec,
       volume = {743},
       number = {1},
          eid = {97},
        pages = {97},
          doi = {10.1088/0004-637X/743/1/97},
archivePrefix = {arXiv},
       eprint = {1111.5599},
 primaryClass = {astro-ph.EP},
       adsurl = {https://ui.adsabs.harvard.edu/abs/2011ApJ...743...97T}
}

@article{Pal2012,
  author  = {{P{\'a}l}, A.},
  title   = {Light-curve modelling for mutual transits},
  journal = {\mnras},
  volume  = {420},
  pages   = {1630},
  year    = {2012},
  doi     = {10.1111/j.1365-2966.2011.20151.x}
}

@article{Short2018,
  author  = {{Short}, D.~R. and {Orosz}, J.~A. and {Windmiller}, G. and {Welsh}, W.~F.},
  title   = {Accurate computation of light curves and the {R}ossiter-{M}c{L}aughlin effect in multibody eclipsing systems},
  journal = {\aj},
  volume  = {156},
  pages   = {297},
  year    = {2018},
  doi     = {10.3847/1538-3881/aae889}
}

@article{HippkeHeller2022,
  author  = {{Hippke}, M. and {Heller}, R.},
  title   = {Pandora: A fast open-source exomoon transit detection algorithm},
  journal = {\aap},
  volume  = {662},
  pages   = {A37},
  year    = {2022},
  doi     = {10.1051/0004-6361/202243129},
  eprint  = {2205.09410}
}

@article{Kalman2023,
  author  = {{K{\'a}lm{\'a}n}, Sz. and {Csizmadia}, Sz. and {Simon}, A.~E. and {Lam}, K.~W.~F. and {Deline}, A. and {Harre}, J.-V. and {Szab{\'o}}, Gy.~M.},
  title   = {Modelling the light curves of transiting exomoons: a zero-order photodynamic agent added to the {T}ransit and {L}ight {C}urve {M}odeller},
  journal = {\mnras},
  volume  = {528},
  pages   = {L66},
  year    = {2024},
  doi     = {10.1093/mnrasl/slad169},
  eprint  = {2311.04647}
}

@article{Csizmadia2020,
  author  = {{Csizmadia}, Sz.},
  title   = {The {T}ransit and {L}ight {C}urve {M}odeller},
  journal = {\mnras},
  volume  = {496},
  pages   = {4442},
  year    = {2020}
}

@article{MandelAgol2002,
  author  = {{Mandel}, K. and {Agol}, E.},
  title   = {Analytic light curves for planetary transit searches},
  journal = {\apjl},
  volume  = {580},
  pages   = {L171},
  year    = {2002}
}

@article{Agol2020,
  author  = {{Agol}, E. and {Luger}, R. and {Foreman-Mackey}, D.},
  title   = {Analytic planetary transit light curves and derivatives for stars with polynomial limb darkening},
  journal = {\aj},
  volume  = {159},
  pages   = {123},
  year    = {2020},
  doi     = {10.3847/1538-3881/ab4fee}
}

@article{Luger2019,
  author  = {{Luger}, R. and {Agol}, E. and {Foreman-Mackey}, D. and {Fleming}, D.~P. and {Lustig-Yaeger}, J. and {Deitrick}, R.},
  title   = {starry: Analytic occultation light curves},
  journal = {\aj},
  volume  = {157},
  pages   = {64},
  year    = {2019},
  doi     = {10.3847/1538-3881/aae8e5}
}

@article{Luger2017,
  author  = {{Luger}, R. and {Lustig-Yaeger}, J. and {Agol}, E.},
  title   = {Planet-planet occultations in {TRAPPIST}-1 and other exoplanet systems},
  journal = {\apj},
  volume  = {851},
  pages   = {94},
  year    = {2017}
}

@article{CabreraSchneider2007,
  author  = {{Cabrera}, J. and {Schneider}, J.},
  title   = {Detecting companions to extrasolar planets using mutual events},
  journal = {\aap},
  volume  = {464},
  pages   = {1133},
  year    = {2007}
}

@article{Librino2014,
  author  = {{Librino}, F. and {Levorato}, M. and {Zorzi}, M.},
  title   = {An algorithmic solution for computing circle intersection areas and its applications to wireless communications},
  journal = {Wireless Communications and Mobile Computing},
  year    = {2014},
  doi     = {10.1002/wcm.2305}
}

@misc{Weisstein,
  author       = {{Weisstein}, E.~W.},
  title        = {Circle-Circle Intersection},
  howpublished = {From MathWorld---A Wolfram Web Resource, \url{https://mathworld.wolfram.com/Circle-CircleIntersection.html}},
  year         = {2024}
}

@misc{Kahan2014,
  author       = {{Kahan}, W.},
  title        = {Miscalculating Area and Angles of a Needle-like Triangle},
  howpublished = {Lecture notes, University of California, Berkeley},
  year         = {2014}
}

@book{Higham2002,
  author    = {{Higham}, N.~J.},
  title     = {Accuracy and Stability of Numerical Algorithms},
  edition   = {2nd},
  publisher = {SIAM},
  address   = {Philadelphia},
  year      = {2002}
}

@inproceedings{Lam2015,
  author    = {{Lam}, S.~K. and {Pitrou}, A. and {Seibert}, S.},
  title     = {Numba: a {LLVM}-based {P}ython {JIT} compiler},
  booktitle = {Proceedings of the Second Workshop on the LLVM Compiler Infrastructure in HPC},
  pages     = {7:1},
  year      = {2015}
}

@article{Harris2020,
  author  = {{Harris}, C.~R. and others},
  title   = {Array programming with {NumPy}},
  journal = {Nature},
  volume  = {585},
  pages   = {357},
  year    = {2020}
}

@article{Hunter2007,
  author  = {{Hunter}, J.~D.},
  title   = {Matplotlib: A {2D} graphics environment},
  journal = {Computing in Science \& Engineering},
  volume  = {9},
  pages   = {90},
  year    = {2007}
}

@manual{mpmath,
  key    = {mpmath},
  author = {Johansson, Fredrik and others},
  title  = {mpmath: a {P}ython library for arbitrary-precision floating-point arithmetic (version 1.4.1)},
  note   = {{\tt https://mpmath.org/}},
  year   = {2026},
}

@misc{eulerr,
  author       = {{Larsson}, J.},
  title        = {eulerr: Area-Proportional {E}uler and {V}enn Diagrams with Ellipses},
  howpublished = {R package version 7.1.0, \url{https://CRAN.R-project.org/package=eulerr}},
  year         = {2026}
}

@misc{vennjs,
  author       = {{Frederickson}, B.},
  title        = {venn.js: area proportional {V}enn and {E}uler diagrams in {J}ava{S}cript},
  howpublished = {\url{https://github.com/benfred/venn.js}, commit 5ff0ef3},
  year         = {2018}
}

@misc{mplvenn,
  author       = {{Tretyakov}, K.},
  title        = {matplotlib-venn: area-weighted {V}enn diagrams for {M}atplotlib},
  howpublished = {version 1.1.2, \url{https://github.com/konstantint/matplotlib-venn}},
  year         = {2025}
}

@misc{shapely,
  author       = {{Gillies}, S. and others},
  title        = {Shapely: manipulation and analysis of geometric objects},
  howpublished = {version 2.1.2, \url{https://github.com/shapely/shapely}},
  year         = {2025}
}

@misc{geos,
  author       = {{GEOS contributors}},
  title        = {{GEOS} computational geometry library},
  howpublished = {version 3.14.1, \url{https://libgeos.org}},
  year         = {2025}
}

\end{document}